# Dwarf novae at low mass transfer rates

J.P. Lasota[1] J.M. Hameury[2] J.M. Huré[3]

[1] UPR 176 du CNRS; DARC, Observatoire de Paris, 92195 Meudon, France
[2] URA 1280 du CNRS, Observatoire de Strasbourg, 11 rue de l'Université, 67000 Strasbourg, France
[3] URA 173 du CNRS; DAEC Observatoire de Paris, 92195 Meudon, France

**Abstract.** We show that if the inner regions of accretion discs in quiescent dwarf nova systems are removed by magnetic disruption or evaporated by siphon flows, the remaining disc is globally stable for mass transfer rates $\lesssim 10^{15}$ g s$^{-1}$. This implies that (super)outbursts in these systems have to be triggered by an enhanced mass transfer form the companion. We suggest that the lack of normal outbursts in WZ Sge results only from its low mass transfer rate and that the viscosity in its disc does not have to be few orders of magnitude lower than in other quiescent dwarf novae.

**Key words:** accretion: accretion disks – stars: cataclysmic variables – stars: individual: WZ Sge

## 1. Introduction

Dwarf novae (DN) are cataclysmic variables (CV) (see e.g. Warner 1995) which, at usually irregular intervals, show a sudden increase of brightness (2 – 7 mag). SU UMa systems undergo rare 'superoutbursts' separated by 'normal' (U Gem type) outbursts. Superoutbursts have larger amplitude than normal outbursts, last longer and show in their light curve, a 'superhump' at a period slightly longer than the orbital one.

In dwarf novae like in other CVs (except in strongly magnetised ones) a white dwarf accretes, through an accretion disc, matter transferred from a Roche lobe filling, lower main sequence companion. There is no doubt that DN outbursts have their origin in the accretion disc but it is not certain that outbursts are pure disc phenomena that occur without some contribution from the companion. The disc instability model (DIM) (see Cannizzo 1993) seems to account for the most general properties of 'normal' outbursts even if it requires some modifications and additions (Livio & Pringle 1992; Meyer & Meyer-Hofmeister 1994)

Send offprint requests to: J.P. Lasota

but these concern the inner disc structure and not the companion's behaviour. In other words, normal outbursts can be considered to occur at a constant mass transfer rate.

In the case of superoutbursts, two kind of models have been considered. On the one hand Osaki (1989) proposed a model in which superoutbursts are due to a combination of thermal and tidal instabilities. On the other hand Osaki (1985), Whitehurst & King (1991), Whitehurst (1994) and Smak (1995) suggested that superoutbursts could be connected to an enhanced mass transfer (EMT) from the companion.

In this letter we show that modifications of the inner disc structure, required for DIM to fit observations, imply that outbursts of dwarf novae at very low mass transfer rate have to be triggered by EMT.

In section 2 we discuss the proposed models of the inner disc and the evidence for increased mass transfer rate before and during outbursts and superoutburst. Section 3 presents the arguments in favour of the presence of a stationary disc during the quiescence of systems with very low transfer rates such as WZ Sge and HT Cas. Conclusions are given in Section 4.

## 2. The Disc instability model versus observations

### 2.1. The UV delay problem

In several dwarf nova systems one observes that the rise to outburst of the optical flux precedes that of the ultraviolet flux. This so-called UV delay is observed both in U Gem and SU UMa systems and ranges from 0.5 to 1 day. The disc instability model in its standard form cannot account for such long delays (see e.g. Pringle et al. 1986; Meyer & Meyer-Hofmeister 1989). When the thermal instability starts in the outer disc regions where most of the optical light is emitted, the transition front which lights up the disc propagates too fast towards the inner UV emitting regions for the resulting spectral behaviour to agree with observations.

are removed from the quiescent cold disc. In this case the heating front would arrive at its usual speed to the inner disc edge but then the 'missing' UV emitting regions would be rebuilt in a viscous time, comparable to the observed UV delay.

A weak magnetic field ($\lesssim 10^5$ G) would disrupt the inner disc at a distance of few to ten white dwarf radii. Livio & Pringle (1992) showed that including a white dwarf magnetic field $\sim 10^4$ G into the standard DIM may account for the observed UV delay.

Meyer & Meyer-Hofmeister (1994) studied the structure of the corona above a quiescent dwarf nova disc. They found that the frictionally heated corona evaporates because of the inefficient radiative cooling of the optically thin gas. As a result a hole forms in the inner disc. A similar idea has been applied to quiescent Soft X-ray Transients (SXT) by Narayan et al. (1995). It is more difficult than in magnetic field case to find a general formula for the inner radius of the evaporated disc but calculations suggest that its value will be also from few to ten white dwarf radii.

2.2. The quiescent UV and X-ray flux problem

A 'hole' in the inner disc would also help to remove another difficulty of the DIM: in some quiescent DN systems one observes an UV and hard X-ray emission that corresponds to accretion at rates higher than those allowed by the model. The DIM requires the surface density of the quiescent disc $\Sigma$ to be less than the critical surface density $\Sigma_{\max}$ everywhere in the disc (see Fig. 1). This implies that an inequality has to be satisfied by the accretion rate in a quiescent disc (Lasota 1995a):

$$\dot M(r) \approx 4.1 \times 10^{14} t_6^{-1} r_9^{3.11} M_1^{-0.37} \left(\frac{\alpha}{0.01}\right)^{-0.79} \quad \text{g s}^{-1} \quad (1)$$

where $\nu$ is the kinematic viscosity coefficient, $\alpha$ the viscosity parameter, $M_1$ the white dwarf mass in solar units, $t_6$ is the recurrence time in units of $10^6$ s and $r_9$ is the radius in $10^9$ cm. In several systems however the observed UV and hard X-ray flux imply accretion rates at the inner disc edge of at least one order of magnitude higher than the one given by Eq. (1) for $r_{\rm in} = r_9 = 1$ (see Meyer & Meyer-Hofmeister 1994). As the maximum allowed accretion rate in a quiescent disc varies as $r^{3.11}$, the removal of the inner disc regions by magnetic stress or evaporation could easily solve the problem of DIM constraints on the quiescent accretion rate without assuming an unusually small viscosity parameter $\alpha$.

Both the UV delay problem and the too high quiescent UV and X-ray flux problem can be overcome by only one effect: the removal of the inner disc in quiescence.

The superhumps observed during SU UMa superoutbursts are due to a tidal distortion of the outer disc resulting from the presence of the 3:1 resonance inside the disc (see e.g. King 1994). It requires mass ratios $q = M_2/M_1 < q_{\rm crit} \approx 0.25 - 0.33$; It is remarkable that U Gem showed a very long (45 day) outburst which looks like a superoutbursts (see Mason et al. 1988) but was not classified as such because a superhump was absent. The mass ratio for this system is $\sim 0.46$.

According to Osaki (1989) superoutbursts are pure disc phenomena. His tidal-thermal instability model makes several predictions that do not seem to be confirmed by observations (Smak 1991, Whitehurst 1994). In addition the thermal-tidal instability model cannot explain the "superoutburst" of U Gem and such very long and rare outbursts in systems with $q > q_{\rm crit}$ would require a different explanation.

A (moderate) mass-transfer enhancement model for superoutbursts was proposed by Whitehurst (1994). Such enhancements are observed in several systems (Smak 1995). EMT would explain the length of the superoutburst. The exact cause of EMT is not known but there is growing evidence that mass transfer rate in CVs is subject to substantial fluctuations on timescales from days to months. In the pure EMT model, however, it is difficult to generate the disc excentricity which is supposed to give rise to the superhump (Lubow 1994). It seems that a 'hybrid' model might be necessary to explain superoutbursts (Smak 1995).

WZ Sge has the particularity of showing only superoutbursts with a recurrence time of $\sim 33$ years and no normal outbursts have been observed in this system. A few other systems show similar but less extreme behaviour. In order to account for such long recurrence time, DIM requires WZ Sge to have an extremely low quiescent viscosity: $\alpha_{\rm cold} < 5 \times 10^{-5}$ (Smak 1993) as compared to the usual value of $\sim 0.01$. The physical reason for such a small value would require an explanation. A similar difficulty is encountered when DIM is applied to SXTs (Lasota 1995a).

## 3. Dwarf novae at very low mass transfer rates

At a given radius $r$, thermal equilibria of accretion discs in cataclysmic variables form, on the $\Sigma$–$\dot M$ (or $\Sigma$–$T_{\rm eff}$) plane, a S–shaped curve, where the middle branch with the negative slope is thermally (and viscously) unstable. Hot stable equilibria exist therefore only for $\Sigma > \Sigma_{\min}$ and cold ones only for $\Sigma < \Sigma_{\max}$ (see e.g. Cannizzo 1993). For the disc to be globally unstable, i.e. unstable at some radius $r_{\rm in} \leq r \leq r_{\rm out}$ one requires the mass transfer rate $\dot M_T$ to satisfy the inequality

$$\dot M \left(\Sigma_{\max}(r_{\rm in})\right) < \dot M_T < \dot M \left(\Sigma_{\min}(r_{\rm out})\right) \quad (2)$$

The range of unstable mass transfer rates therefore depends on the size of the disc. Fig. 1 shows surface density

$M_1$ (Hameury et al. 1995). The accretion rate $\dot M$ is constant for each curve.

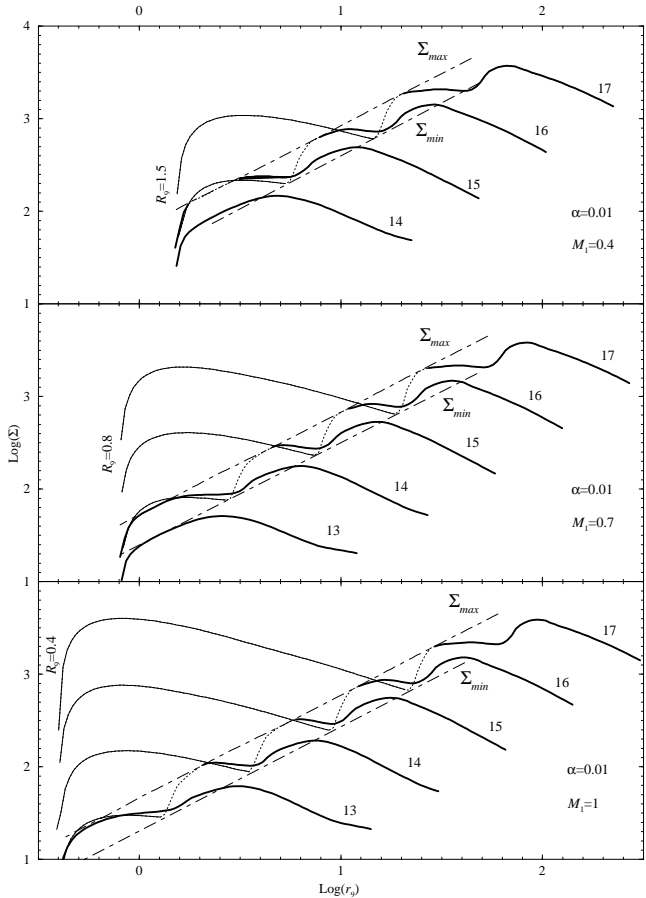

**Fig. 1.** Density profiles of stationary accretion discs around white dwarfs with $M_1 = 0.4$ (upper panel), $M_1 = 0.7$ (middle panel) and $M_1 = 1$ (lower panel), for $\dot M = 10^{13}$, $10^{14}$, $10^{15}$, $10^{16}$ and $10^{17}$ g s$^{-1}$ (except for $M_1 = 0.4$ for which the $10^{13}$ profile is not plotted). The viscosity parameter $\alpha$ is equal to 0.01. The stable parts of *cold* equilibria are represented by a thick line, the unstable part by a dotted line and the hot stable equilibria by a continous thin line. For $M_1 = 0.4$ the $10^{15}$ g s$^{-1}$ profile is represented by a dashed-dotted line when it coincides with $\Sigma_{\max}$ i.e. when the disc is marginally stable.

The part of the curve which is above the $\Sigma_{\max}$ line cannot represent a cold thermal equilibrium so that the disc cannot be globally in equilibrium ($\dot M = const$) for the corresponding mass transfer rate.

From Fig. 1 one can see that there exist two types of equilibria represented by the $\Sigma(r)$ curves: for $\dot M \lesssim \dot M'$ (where $\dot M'$ is some value of accretion rate depending on $M_1$ and $\alpha$; in the Fig. 1 $\dot M' \sim 10^{13} - 10^{14}$ g s$^{-1}$) the surface density is always less than $\Sigma_{\max}$. For such low $\dot M$ the disc is globally in cold stable equilibrium, from $r_{\text{out}}$ down part of the disc can be in a cold stable equilibrium. Starting at some $r_{\text{out}}$ from $\Sigma < \Sigma_{\max}$, the $\Sigma(r)$ curve reaches, with decreasing $r$, the value $\Sigma_{\max}$ at some radius $r_{\text{crit}}$. The disc can be in cold equilibrium only for $r_{\text{crit}} < r < r_{\text{out}}$. The next segment of the curve between $\Sigma_{\max}$ and $\Sigma_{\min}$ is unstable while the curve to the left of $\Sigma_{\min}$ represent hot equilibria. For $\dot M > \dot M'$ the disc is therefore globally unstable if it extends down to the surface of the white dwarf, and a cold steady state cannot exist. There exists however a radius $r_{\text{in}} = r_{\text{crit}}$ for which is disc is globally stable at a $\dot M = const$. As mentioned above both the magnetically disrupted and the evaporated accretion discs have inner radii $\approx 3 \times 10^9$ cm. Fig. 2 shows $\dot M (\Sigma_{\max}(r)) \equiv \dot M_{\text{crit}}(r)$. One can see that for $r_{\text{in}} \sim 2.5 - 5 \times 10^9$ (depending on $M_1$) the critical mass transfer rate below which the disc is globally stable is $\dot M_{\text{crit}} \approx 10^{15}$ g s$^{-1}$.

If models assuming a 'hole' in the inner disc were to apply to CVs transferring mass at low rates, those systems in quiescence would be globally stable and outbursts could only be triggered by an *increase of mass transfer rate* form the companion. One should stress here that the resulting outburst would be due to the 'usual' thermal instability, as the only effect of the enhanced mass transfer is to bring the disc into a globally unstable state (and to add some mass into the disc). The proposed model scheme is therefore a 'hybrid' (EMT-DIM) model.

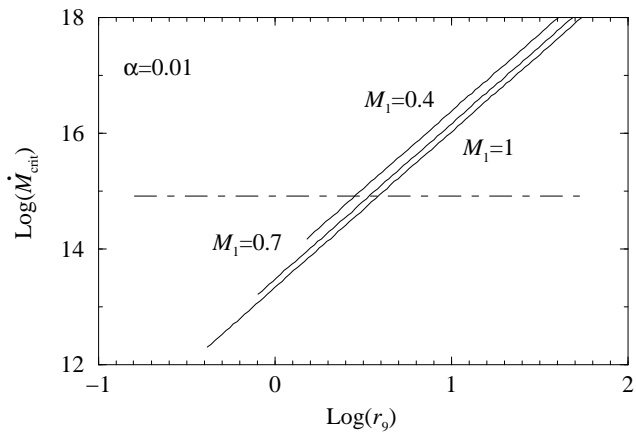

**Fig. 2.** $\dot M (\Sigma_{\max}(r)) \equiv \dot M_{\text{crit}}(r)$ for three values of $M_1 = 0.4$, 0.7 and 1. At a given $r'$, mass transfer rates $\dot M < \dot M_{\text{crit}}(r')$ correspond to global, stable disc equilibria for $r > r'$.

There are no reasons for the disrupted or evaporated disc models not to apply to systems at low mass transfer rates. On the contrary, both types of models were devised to explain the behaviour of discs at low accretion rates.

One of such systems is WZ Sge for which Smak (1993) estimated the mass transfer rate to be $\sim 10^{15}$ g s$^{-1}$ and the white dwarf mass $M_1 = 0.4$. Fig. 2 shows that at this ally stable even a very weak magnetic field would make it stable. This system should not show dwarf nova outbursts. Indeed, WZ Sge shows only superoutbursts with a very long recurrence time. Those superoutbursts would be due to a thermal instability triggered by EMT from the companion. It is interesting to note that Patterson (1994) suggested that the white dwarf in WZ Sge is weakly magnetized.

Our model allows a unified scheme for SU UMa's and WZ Sge type systems. If SU UMa's superoutbursts are triggered by enhanced mass transfer, the difference between those systems and WZ Sge and similar binaries would be due only to the value of the mass transfer from the companion. In SU UMa's mass transfer rates are close to $10^{16}$ g s$^{-1}$ and $\dot M_T$ can be even larger than $\dot M_{\rm crit}$ ($r_{\rm crit} \gtrsim r_{\rm out}$) so that, contrary to WZ Sge, their discs are globally unstable for any value of the inner radius and they undergo 'normal' outbursts and superoutbursts are triggered by EMT. There is no need in the proposed scheme to assume, for unknown reasons, extremely low values of $\alpha$ in WZ Sge. True, our scheme involves mass transfer modulations whose reasons are also unknown but at least such modulations are observed in many systems. One possible reason for those modulations could be appearance of magnetic spots at the inner Lagrange point (Livio & Pringle 1994).

Of special interest is the system HT Cas which show rare outbursts (Wenzel 1987) of which at least one was a superoutburst (Zhang et al. 1986). According to Wood et al. (1992) the mass transfer rate in this system is $\sim 2 \times 10^{15}$ g s$^{-1}$. Eclipse mapping shows a very flat slope of the temperature in the inner disc. This could be due to magnetic disruption and/or to nonradiative energy loss (see e.g. Rutten et al. 1992). HT Cas was also observed in a 'low state' with accretion (and mass transfer) rate of $\sim 6 \times 10^{13}$ g s$^{-1}$ (Wood et al. 1995). We suggest that the strange 'erratic' behaviour of HT Cas is due to the fact that its mass transfer rate is very close to $\dot M_{\rm crit}$ and quite often the disc is globally stable. In any case the mass transfer rate in this system *is observed* to vary on short time scales.

As shown by Narayan et al. (1995) quiescent spectra of soft X-ray transients can be explained by advection dominated flows near the central black hole. A scheme similar to the one proposed here may be used to explain SXTs outbursts (Yi et al. 1995, Lasota 1995b).

## 4. Conclusions

We have shown that the current ideas on the inner accretion disc structure of quiescent dwarf novae imply that CVs with very low mass transfer rates from the companion are globally stable with respect to the thermal instability. Thus, outbursts can occur only during episodes of enhanced mass transfer.

no (or only rare) 'normal' outbursts. The viscosity in the discs of those systems would be comparable to the one in binaries with slightly higher mass transfer rate. The proposed scheme naturally fits into the models in which superoutbursts are triggered by EMT.